\documentclass[prd,preprint,preprintnumbers,nofootinbib,eqsecnum,superscriptaddress]{revtex4}

 \usepackage[dvips,final]{graphicx}
  \usepackage{amssymb}
   \usepackage{amsmath}
    \usepackage{amsfonts}
     \usepackage{epsfig}
      \usepackage{bm}

\usepackage{mathpazo}

\usepackage[section]{placeins}

\input epsf.tex
\def\desepsf(#1 width #2){\epsfxsize=#2 \epsfbox{#1}}

\usepackage[normalem]{ulem}

\usepackage{multirow}
\usepackage{ctable}
\usepackage{booktabs}
\usepackage{array}
\usepackage{tabularx}
\usepackage{xcolor}
\usepackage{pstricks}
\definecolor{fred}{rgb}{0.90053, 0.00369, 0.00159}  

\newcommand{\be}{\begin{eqnarray}}
\newcommand{\ee}{\end{eqnarray}}

\begin{document}

\author{Rafa{\l} Maciu{\l}a}
\email{rafal.maciula@ifj.edu.pl}
\affiliation{Institute of Nuclear
Physics, Polish Academy of Sciences, ul. Radzikowskiego 152, PL-31-342 Krak{\'o}w, Poland}

\author{Antoni Szczurek}
\email{antoni.szczurek@ifj.edu.pl} 
\affiliation{Institute of Nuclear Physics, Polish Academy of Sciences, ul. Radzikowskiego 152, PL-31-342 Krak{\'o}w, Poland}
\affiliation{College of Natural Sciences, Institute of Physics, University of Rzesz{\'o}w, ul. Pigonia 1, PL-35-310 Rzesz{\'o}w, Poland}



\title{Recombination mechanism for $\bm{D^{0}}$-meson production and $\bm{D^{0}\!-\!\overline{D^{0}}}$ production asymmetry in the LHCb $\bm{p\!+\!\!^{20}\!N\!e}$ fixed-target experiment}

\begin{abstract}
We discuss production of neutral $D$ mesons 
in proton-proton collisions at the LHC
(fixed target mode) in the framework of the BJM recombination model.
We present rapidity and transverse momentum distributions of 
$D$ mesons and compare the recombination contribution 
to the dominant gluon-gluon fusion mechanism.
Both the direct production, as dictated by the matrix element, 
and fragmentation of the associated $c$ or $\bar c$ are included. 
The latter mechanism generates $D$ mesons with smaller rapidities than 
those produced directly.
We calculate the $D^0 + {\overline D^{0}}$ meson distributions relevant for
fixed target $p\!+\!\!^{4}\!H\!e$ collisions at $\sqrt{s}$ = 86.6 GeV as well as for $p\!+\!\!^{20}\!N\!e$ collisions at $\sqrt{s}$ = 69 GeV.
The recombination component is consistent with the LHCb data and in addition results in production asymmetry.
The asymmetries in $D^{0}\!-\!\overline{D^{0}}$ production as a function 
of rapidity and transverse momentum are shown and the cancellation 
of terms for direct production and associated $c/{\bar c}$
  fragmentation is discussed. 
\end{abstract} 

\maketitle

\section{Introduction}

The production mechanism of midrapidity $D$ mesons is relatively well
known. At low energy the quark-antiquark annihilation and gluon-gluon
fusion must be included in the leading order calculation.
At higher energy the gluon-gluon fusion is the dominant mechanism. 
We calculate the latter contribution in the $k_T$-factorization
approach. As already discussed in the literaturte such an approach leads 
to a good description of single $D$ meson production data at high
energies, including RHIC \cite{Maciula:2015kea} and 
the LHC \cite{Maciula:2013wg,Maciula:2019izq}, as well as 
of correlation observables for $D {\bar D}$-pair \cite{Maciula:2013wg,Karpishkov:2016hnx} and $DD$-pair production \cite{Maciula:2013kd,vanHameren:2014ava}. Recently, it has been also applied for open charm meson production at low energies within the fixed-target mode of the LHCb experiment \cite{Maciula:2020cfy,Maciula:2021orz}. 

On the other hand the mechanism of forward/backward production
of $D$ mesons is rather poorly understood. Two mechanisms are usually 
considered in this context in phenomenological applications:
(a) mechanism related to the knock-out of preexisting 
intrinsic $c / \bar c$ from the nucleon \cite{BHPS1980,Maciula:2020dxv,Maciula:2021orz}; 
(b) mechanism of recombination of $q/ \bar q$ with $\bar c / c$
    as proposed by Braaten-Jia-Mechen \cite{BJM2002a,BJM2002b,BJM2002c}.
We shall call the first mechanism as intrinsic charm (IC) mechanism 
and the second as BJM recombination.

As discussed e.g. in Refs.~\cite{Maciula:2020dxv,Maciula:2021orz} the Brodsky et al. approach to 
the intrinsic charm \cite{BHPS1980} may lead to the production of 
large rapidity $D$ mesons and consequently to large rapidity 
neutrino/antineutrino production \cite{Maciula:2020dxv}.
The BHPS IC was included recently into a global parton analysis
\cite{Hou:2017khm}.
The original BHPS formulation leads to $c_{IC}(x) = \bar c_{IC}(x)$. 
This mechanism was discussed recently in the context
of fixed target experiments \cite{Maciula:2021orz} as well as 
of high-energy (anti)neutrinos observed recently by the IceCube 
laboratory \cite{Goncalves:2021yvw} at the South Pole.
There only upper limit for the BHPS intrinsic charm can be obtained
\cite{Goncalves:2021yvw} as there can be extragalatic sources of very high 
(anti)neutrinos \cite{IceCube:2020wum}.
Another mechanism of internal charm-anticharm component in the nucleon 
is related to the meson cloud model \cite{Melnitchouk:1997ig,Cazaroto:2013wy}.
In contrast to the BHPS model it generates asymmetric intrinsic 
$c/{\bar c}$ distribution in the nucleon. 
Similar model with light meson-baryon Fock components in the nucleon 
was able to explain $\bar d - \bar u$ asymmetry in the nucleon, see e.g.
\cite{Holtmann:1996be}, not possible to be explained in perturbative calculations.
As will be discussed here the $\bar d - \bar u$ asymmetry in the nucleon,
probably of the meson cloud origin, may have some influence on
$D - \bar D$ production asymmetries obtained in the recombination model.
In our calculations here we shall include distributions with
${\bar d}-{\bar u}$ asymmetry.

Here we rather concentrate on the recombination mechanism
which may be considered as an unwanted background for 
the large-$x$ intrinsic charm contribution 
(which we will call here BHPS model for brevity).
In this mechanism there are two-body final states such as 
$D_i$ and $c$ or ${\bar D}_i$ and $\bar c$.
We shall consider production of different pseudoscalar and vector $D$
mesons as well as feed down due to decays of vector $D^*$ mesons into
measured pseudoscalar $D$ mesons.
Somewhat different model was used to explain asymmetries
in $D$ meson production in \cite{Rapp:2003wn}.

The latter approach is based on the formalism proposed long ago 
by Das and Hwa \cite{Das:1977cp}. In this approach the transverse momentum
of $D$ mesons is not explicit in contrast to the BJM model.
Stll another recombination approach was presented
in \cite{Berezhnoy:2000ji}, but in the context of photoproduction. 
All the previous studies of $D$ meson recombination concentrated on
low-energy $p p$, $\pi^- p$ or $\gamma p$ processes.
According to our knowledge recombination was not studied for higher energies.
In Ref.~\cite{Maciula:2017wov} a subleading fragmentation
${\bar d} \to D^+, d \to D^-$  was discussed to explain observed 
asymmetry in $D^+ / D^-$ production as observed in 
the LHCb experiment \cite{LHCb:2012fb}.

In principle, the asymmetry in $D$ meson production can be obtained also
in the Lund string approach \cite{Norrbin:2000zc}. However, we shall not
discuss it in the present paper.

In the present paper we wish to estimate the contribution
of the BJM mechanism for fixed target experiment(s) at the LHC.

\section{Details of the model calculations}

In the present study we take into consideration three different
production mechanisms of charm, including: a) the standard (and usually
considered as a leading) QCD mechanism of gluon-gluon fusion: $gg \to
c\bar c$ ; b) the mechanism driven by the intrinsic charm component of
proton: $gc \to gc$; and c) the recombination mechanism: $gq \to
Dc$. Calculations of the first two contributions are performed following
our previous study in Ref.~\cite{Maciula:2021orz}. Some of corresponding
histograms shown there are repeated here. What is completely new in the
present study is the BJM recombination component which was not considered in this context in the past.         

\subsection{The standard QCD mechanism for charm production}

\begin{figure}[!h]
\centering
\begin{minipage}{0.3\textwidth}
  \centerline{\includegraphics[width=1.0\textwidth]{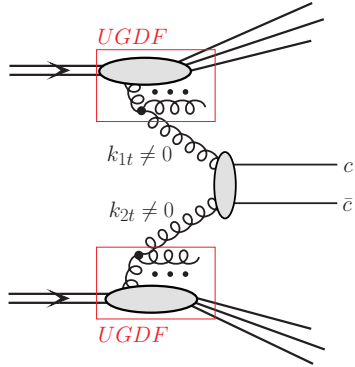}}
\end{minipage}
  \caption{
\small A diagram of the standard QCD mechanism of charm production in the $k_{T}$-factorization approach driven by the fusion of two off-shell gluons.
}
\label{fig:diagramLO}
\end{figure}

As was already mentioned here we follow the theoretical formalism for
the calculation of the $c\bar{c}$-pair production in the
$k_{T}$-factorization approach \cite{kTfactorization}, as adopted and
discussed in the context of the LHCb fixed-target charm data in
Ref.~\cite{Maciula:2020cfy}. In this framework the transverse momenta
$k_{t}$'s (or virtualities) of both partons entering the hard process are taken into account, both in the matrix elements and in the parton distribution functions. Emission of the initial state partons is encoded in the transverse-momentum-dependent (unintegrated) PDFs (uPDFs). In the case of charm flavour production the parton-level cross section is usually calculated via the $2\to 2$ leading-order $g^*g^* \to c\bar c$ fusion mechanism with off-shell initial state gluons that is the dominant process at high energies (see Fig.~\ref{fig:diagramLO}). Even at lower energies as long as small transverse momenta and not extremely backward/forward rapidities are considered the $q^*\bar q^* \to c\bar c $ mechanism remains subleading. Then the hadron-level differential cross section for the $c \bar c$-pair production, formally at leading-order, reads:
\begin{eqnarray}\label{LO_kt-factorization} 
\frac{d \sigma(p p \to c \bar c \, X)}{d y_1 d y_2 d^2p_{1,t} d^2p_{2,t}} &=&
\int \frac{d^2 k_{1,t}}{\pi} \frac{d^2 k_{2,t}}{\pi}
\frac{1}{16 \pi^2 (x_1 x_2 s)^2} \; \overline{ | {\cal M}^{\mathrm{off-shell}}_{g^* g^* \to c \bar c} |^2}
 \\  
&& \times  \; \delta^{2} \left( \vec{k}_{1,t} + \vec{k}_{2,t} 
                 - \vec{p}_{1,t} - \vec{p}_{2,t} \right) \;
{\cal F}_g(x_1,k_{1,t}^2,\mu_{F}^2) \; {\cal F}_g(x_2,k_{2,t}^2,\mu_{F}^2) \; \nonumber ,   
\end{eqnarray}
where ${\cal F}_g(x_1,k_{1,t}^2,\mu_{F}^2)$ and ${\cal F}_g(x_2,k_{2,t}^2,\mu_{F}^2)$
are the gluon uPDFs for both colliding hadrons and ${\cal M}^{\mathrm{off-shell}}_{g^* g^* \to c \bar c}$ is the off-shell matrix element for the hard subprocess.
The gluon uPDF depends on gluon longitudinal momentum fraction $x$, transverse momentum
squared $k_t^2$ of the gluons entering the hard process, and in general also on a (factorization) scale of the hard process $\mu_{F}^2$. They must be evaluated at longitudinal momentum fractions 
$x_1 = \frac{m_{1,t}}{\sqrt{s}}\exp( y_1) + \frac{m_{2,t}}{\sqrt{s}}\exp( y_2)$, and $x_2 = \frac{m_{1,t}}{\sqrt{s}}\exp(-y_1) + \frac{m_{2,t}}{\sqrt{s}}\exp(-y_2)$, where $m_{i,t} = \sqrt{p_{i,t}^2 + m_c^2}$ is the quark/antiquark transverse mass. 

As we have carefully discussed in Ref.~\cite{Maciula:2019izq}, there is a direct relation between a resummation present in uPDFs in the transverse momentum dependent factorization and a parton shower in the collinear framework. In most uPDF cases the off-shell gluon can be produced either from gluon or quark, therefore, in the $k_{T}$-factorization all channels driven by $gg, q\bar q$ and even by $qg$ initial states are open already at leading-order (in contrast to the collinear factorization).

The kinematical configuration of the fixed-target LHCb experiment corresponds to the region where in principle the Catani-Ciafaloni-Fiorani-Marchesini (CCFM) \cite{CCFM} evolution equation is legitimate for any pQCD theoretical calculations and could, in principle, be used to describe the dynamics behind the mechanisms of \textit{e.g.} open charm meson production. In the numerical calculations below we follow the conclusions from Refs.~\cite{Maciula:2020cfy,Maciula:2021orz} and apply the JH-2013-set2 gluon uPDFs \cite{Hautmann:2013tba} that are determined from high-precision DIS measurements. As a default set in the numerical calculations we take the renormalization scale
$\mu^2 = \mu_{R}^{2} = \sum_{i=1}^{n} \frac{m^{2}_{it}}{n}$ (averaged
transverse mass of the given final state) and the charm quark mass
$m_{c}=1.5$ GeV. The strong-coupling constant $\alpha_{s}(\mu_{R}^{2})$
at next-to-next-to-leading-order is taken from the CT14nnloIC PDF
routines. The CCFM uPDFs here are taken at a rather untypical value of the factorization scale $\mu_{F}^2 = M_{c\bar c}^2 + P_{T}^{2}  $, where $M_{c\bar c}$ and $P_{T}$ are the $c\bar c$-invariant mass (or energy of the scattering subprocess) and the transverse momentum of $c\bar c$-pair (or the incoming off-shell gluon pair). This unusual definition has to be applied as a consequence of the CCFM evolution algorithm \cite{Hautmann:2013tba}.

\subsection{The Intrinsic Charm induced component}

\begin{figure}[!h]
\centering
\begin{minipage}{0.3\textwidth}
  \centerline{\includegraphics[width=1.0\textwidth]{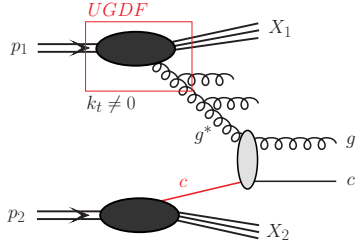}}
\end{minipage}
  \caption{
\small A diagrammatic representation of the intrinsic charm mechanism of charm production within the hybrid model with the off-shell gluon and the on-shell charm quark in the initial state.
}
\label{fig:diagramIC}
\end{figure}

The intrinsic charm contribution to charm production cross section (see
Fig.~\ref{fig:diagramIC}) is obtained within the hybrid theoretical
model discussed by us in detail in Ref.~\cite{Maciula:2020dxv}. The LHCb
fixed-target configuration allows to explore the charm cross section in
the backward rapidity direction where an asymmetric kinematical
configurations are selected. Thus in the basic $gc \to gc$ reaction the gluon PDF and the intrinsic charm PDF are simultaneously probed at different longitudinal momentum fractions - rather intermediate for the gluon and large for the charm quark.

Within the asymmetric kinematic configuration $x_1 \ll x_2$ the cross section for the processes under consideration can be calculated in the so-called hybrid factorization model motivated by the work in Ref.~\cite{Deak:2009xt}. In this framework the small- or intermediate-$x$ gluon is taken to be off mass shell and 
the differential cross section e.g. for $pp \to g c X$ via $g^* c \to g c$ mechanism reads:
\begin{eqnarray}
d \sigma_{pp \to gc X} = \int d^ 2 k_{t} \int \frac{dx_1}{x_1} \int dx_2 \;
{\cal F}_{g^{*}}(x_1, k_{t}^{2}, \mu^2) \; c(x_2, \mu^2) \; d\hat{\sigma}_{g^{*}c \to gc} \; ,
\end{eqnarray}
where ${\cal F}_{g^{*}}(x_1, k_{t}^{2}, \mu^2)$ is the unintegrated
gluon distribution in one proton and $c(x_2, \mu^2)$ a collinear PDF in
the second one. The $d\hat{\sigma}_{g^{*}c \to gc}$ is the hard partonic
cross section obtained from a gauge invariant tree-level off-shell
amplitude. A derivation of the hybrid factorization from the dilute
limit of the Color Glass Condensate approach can be found e.g. in Ref.~\cite{Dumitru:2005gt} (see also Ref.~\cite{Kotko:2015ura}). The relevant cross sections are calculated with the help of the \textsc{KaTie} Monte Carlo generator \cite{vanHameren:2016kkz}. There the initial state quarks (including heavy quarks) can be treated as a massless partons only.  

Working with minijets (jets with transverse momentum of the order of a few GeV) requires a phenomenologically motivated regularization of the cross sections. Here we follow the minijet model \cite{Sjostrand:1987su} adopted e.g. in \textsc{Pythia} Monte Carlo generator, where a special suppression factor is introduced at the cross section level \cite{Sjostrand:2014zea}:
\begin{equation}
F(p_t) = \frac{p_t^2}{ p_{T0}^2 + p_t^2 } \; 
\label{Phytia_formfactor}
\end{equation}
for each of the outgoing massless partons with transverse momentum $p_t$, where $p_{T0}$ is a free parameter of the form factor
that also enters as an argument of the strong coupling constant $\alpha_{S}(p_{T0}^2+\mu_{R}^{2})$.

This suppression factor was originally proposed to remove singularity of minijet cross sections in the collinear approach at leading-order. In the hybrid model (or in the $k_{T}$-factorization) the leading-order cross sections are finite as long as $k_{T}> 0$, where $k_{T}$ is the transverse momentum of the incident off-shell parton. Within this approach, a treatment of the small-$k_{T}$ region in the construction of a given unintegrated parton density is crucial. Different models of uPDFs may lead to different behaviour of the cross section at small minijet transverse momenta but in any case the cross sections should be finite. However, as it was shown in Ref.~\cite{Kotko:2016lej}, the internal $k_{T}$ cannot give a minijet suppression consistent with the minijet model and the related regularization seems to be necessary even in this framework.


In the numerical calculations below, the intrinsic charm PDFs are taken at the initial scale $m_{c} = 1.3$ GeV, so the perturbative charm contribution is intentionally not taken into account. In the numerical calculations below we apply the grid of the intrinsic charm distribution from the CT14nnloIC PDF \cite{Hou:2017khm} that corresponds to the BHPS model \cite{BHPS1980}.    

\subsection{Recombination model and charm production}

\begin{figure}[!h]
\includegraphics[width=5cm]{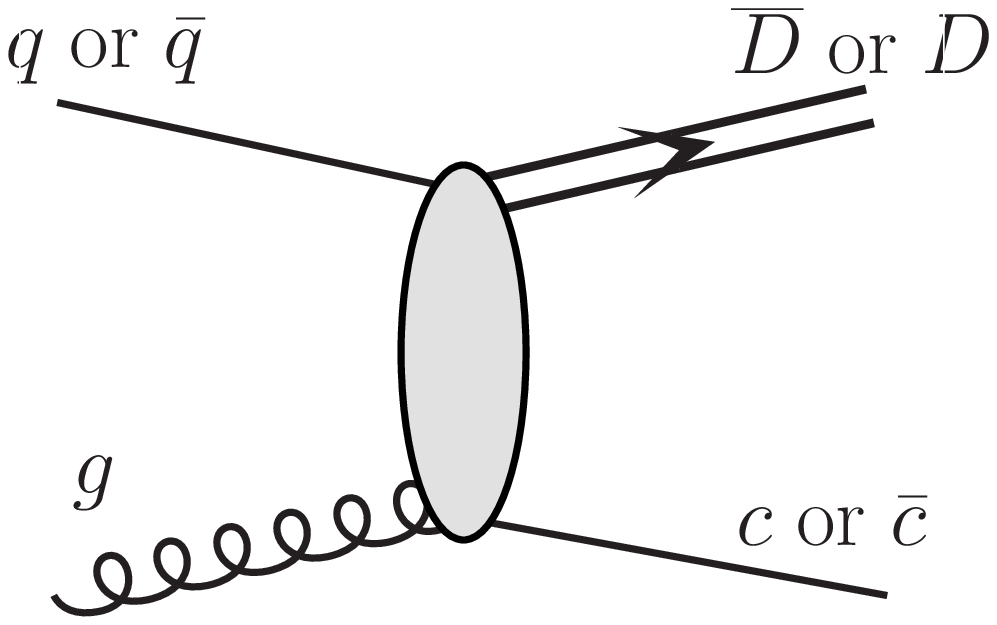}
\hskip+5mm
\includegraphics[width=5cm]{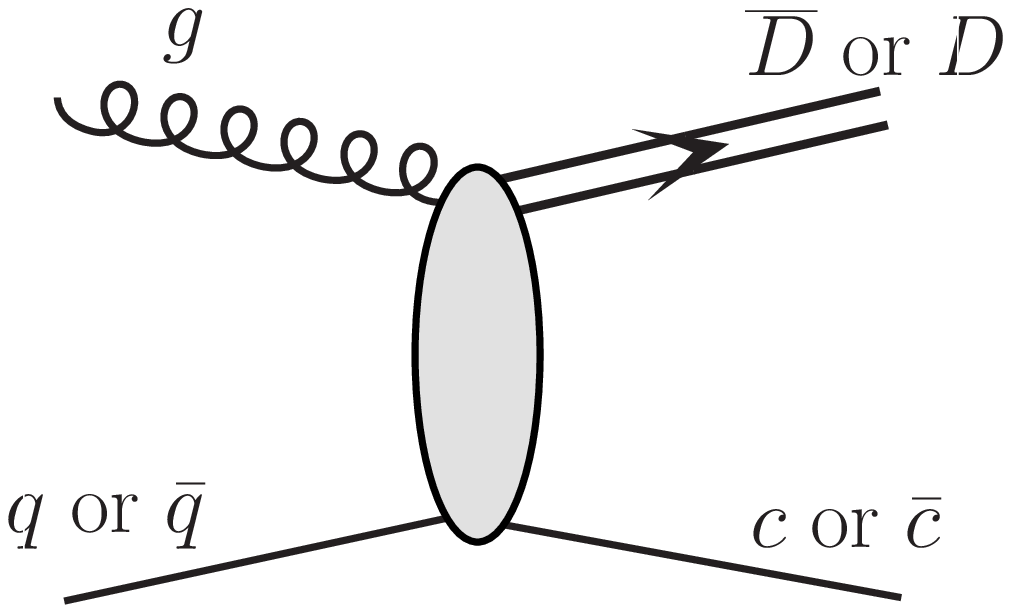}
\caption{Generic leading-order diagrams for $D$ meson production via 
the BJM recombination model.}
\label{fig:recombination_diagrams}
\end{figure}

The underlying mechanism of the BJM recombination is illustrated 
in Fig.~\ref{fig:recombination_diagrams}. Differential cross section for production of $D c$ final state reads:
\begin{eqnarray}
\frac{d\sigma}{d y_1 d y_2 d^2 p_{t}} = \frac{1}{16 \pi^2 {\hat s}^2}
&& [ x_1 q_1(x_1,\mu^2) \, x_2 g_2(x_2,\mu^2)
\overline{ | {\cal M}_{q g \to D c}(s,t,u)|^2} \nonumber \\
&+& x_1 g_1(x_1,\mu^2) \, x_2 q_2(x_2,\mu^2)
\overline{ | {\cal M}_{g q \to D c}(s,t,u)|^2} ]  \, .
\label{cross_section}
\end{eqnarray}
Above $y_1$ is rapidity of the $D$ meson and $y_2$ rapidity of 
the associated $c$ or $\bar c$. The fragmentation of the latter
will be discussed below.

The matrix element squared in (\ref{cross_section}) can be writted as
\begin{equation}
\overline{ | {\cal M}_{q g \to D c}(s,t,u)|^2} =
\overline{ | {\cal M}_{q g \to ({\bar c} q)^n c} |^2} \cdot \rho \; ,
\label{rho_definition}
\end{equation}
where $n$ enumerates quantum numbers of the ${\bar c} q$ system
$n \equiv ^{2J+1}L$.
$\rho$ can be interpreted as a probability to form real meson.
For illustration as our default set we shall take $\rho$ = 0.1, but the precise number
should be adjusted to experimental data. For the discussion of
the parameter see e.g. Refs.~\cite{BJM2002b,BJM2002c} and references therein.
The asymmetries observed in photoproduction can be explained with $\rho$ = 0.15
\cite{BJM2002c}.

The explicit form of the matrix element squared can be found in
\cite{BJM2002a} for pseudoscalar and vector meson production for color
singlet and color octet meson-like states.

Similar formula can be written for production of ${\bar D} {\bar c}$.
Then the quark distribution is replaced by the antiquark distribution.

In the following we include only color singlet $(q {\bar c})^n$
or $({\bar q} c)^n$ components.

The factorization scale in the calculation is taken as:
\begin{equation}
\mu^2 = p_t^2 + \frac{m_{t,D}^2 + m_{t,c}^2}{2} \; .
\label{scale}
\end{equation}

Within the recombination mechanism we include fragmentation of $c$-quarks or $\bar{c}$-antiquarks accompanying directly produced
$D$-mesons or $\overline{D}$-antimesons, e.g.:
\begin{eqnarray}
d \sigma [q g \to \overline{D}_{\mathrm{direct}} + D_{\mathrm{frag.}}] &=& d \sigma [q g \to \overline{D} + c] 
\otimes F^{\mathrm{frag.}}_{c \to D} \; ,
\label{fragmentation}
\end{eqnarray}
where $F^{\mathrm{frag.}}_{c \to D}$ is the relevant fragmentation function.
How the convolution $\otimes$ is understood
is explained in \cite{Maciula:2019iak}.

In explicit calculations of $D$ meson production discussed in the next
section we shall consider proton - isoscalar nucleus scattering.
For relatively light nuclei as $^{20}\!N\!e$ the nuclear effects can be
neglected. Then the $p\!+\!\!^{20}\!N\!e$ scattering is a superposition of
$p+p$ and $p+n$ scatterings.
The quark/antiquark distributions in the neutron are obtained
from those in the proton by assuming isospin symmetry.

We shall discuss in the present paper also the asymmetry in production of
$D^0$ meson and ${\overline D}^0$ antimeson.
The asymmetry is defined as:
\begin{equation}
A_{p} = \frac{d\sigma^{D^0}\!\!/d\xi - d\sigma^{\overline{D}^0}\!\!/d\xi}
                   {d\sigma^{D^0}\!\!/d\xi + d\sigma^{\overline{D}^0}\!\!/d\xi}
\; ,
\label{asymmetry}
\end{equation}
where $\xi$ represents single variable ($y$ or $p_t$) or even a pair of
variables ($(y, p_t)$).

Only a part of the pseudoscalar $D$ mesons is directly produced.
A second part originates from vector meson decays.
The vector $D$ mesons promptly and dominantly decay to pseudoscalar mesons:
\begin{eqnarray}
D^{*0} &\to& D^0 \pi (0.619), D^0 \gamma (0.381) , \\
D^{*+} &\to& D^0 \pi^+ (0.677), D^+ \pi^0 (0.307), D^+ \gamma (0.0016) , \\
D_s^{*+} &\to& D_s^+ \gamma (0.935), D_s^+ \pi^0 (0.058), D_s^+ e^+ e^- (6.7 10^{-3}),  \\
\end{eqnarray}
%

%
%

\subsection{Hadronization of charm quarks}

The transition of charm quarks to open charm mesons is done in the framework of the independent parton fragmentation picture (see \textit{e.g.} Refs.~\cite{Maciula:2019iak}) where the inclusive distributions of open charm meson can be obtained through a convolution of inclusive distributions of charm quarks/antiquarks and $c \to D$ fragmentation functions. Here we follow exactly the method devoted to rather low energies, which was applied by us in our previous study reported in Ref.~\cite{Maciula:2021orz}. According to this approach we assume that the $D$-meson is emitted in the direction of parent $c$-quark/antiquark, i.e. $\eta_D=\eta_c$ (the same pseudorapidities or polar angles) and the $z$-scaling variable is defined with the light-cone momentum i.e. $p^{+}_{c} = \frac{p^{+}_{D}}{z}$ where $p^{+} = E + p$.
In numerical calculations we take the Peterson fragmentation function \cite{Peterson:1982ak} with $\varepsilon = 0.05$, often used in the context of hadronization of heavy flavours. Then, the hadronic cross section is normalized by the relevant charm fragmentation fractions for a given type of $D$ meson \cite{Lisovyi:2015uqa}.  
In the numerical calculations below for $c \to D^{0}$ meson transition we take the fragmentation probability $\mathrm{P}_{c \to D} = 61\%$.

\section{Numerical results}

\begin{figure}[!h]
\begin{minipage}{0.45\textwidth}
  \centerline{\includegraphics[width=1.0\textwidth]{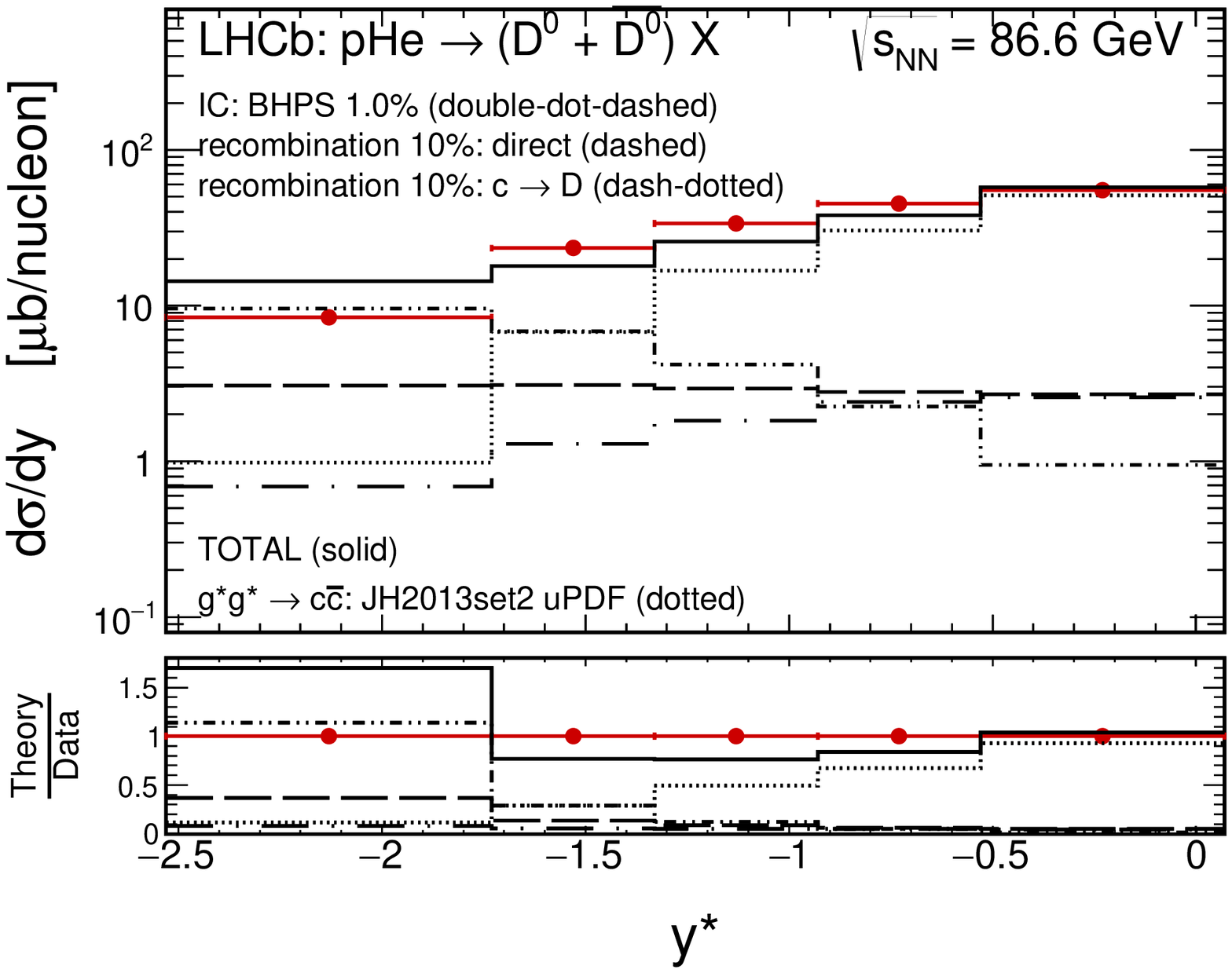}}
\end{minipage}
\begin{minipage}{0.45\textwidth}
  \centerline{\includegraphics[width=1.0\textwidth]{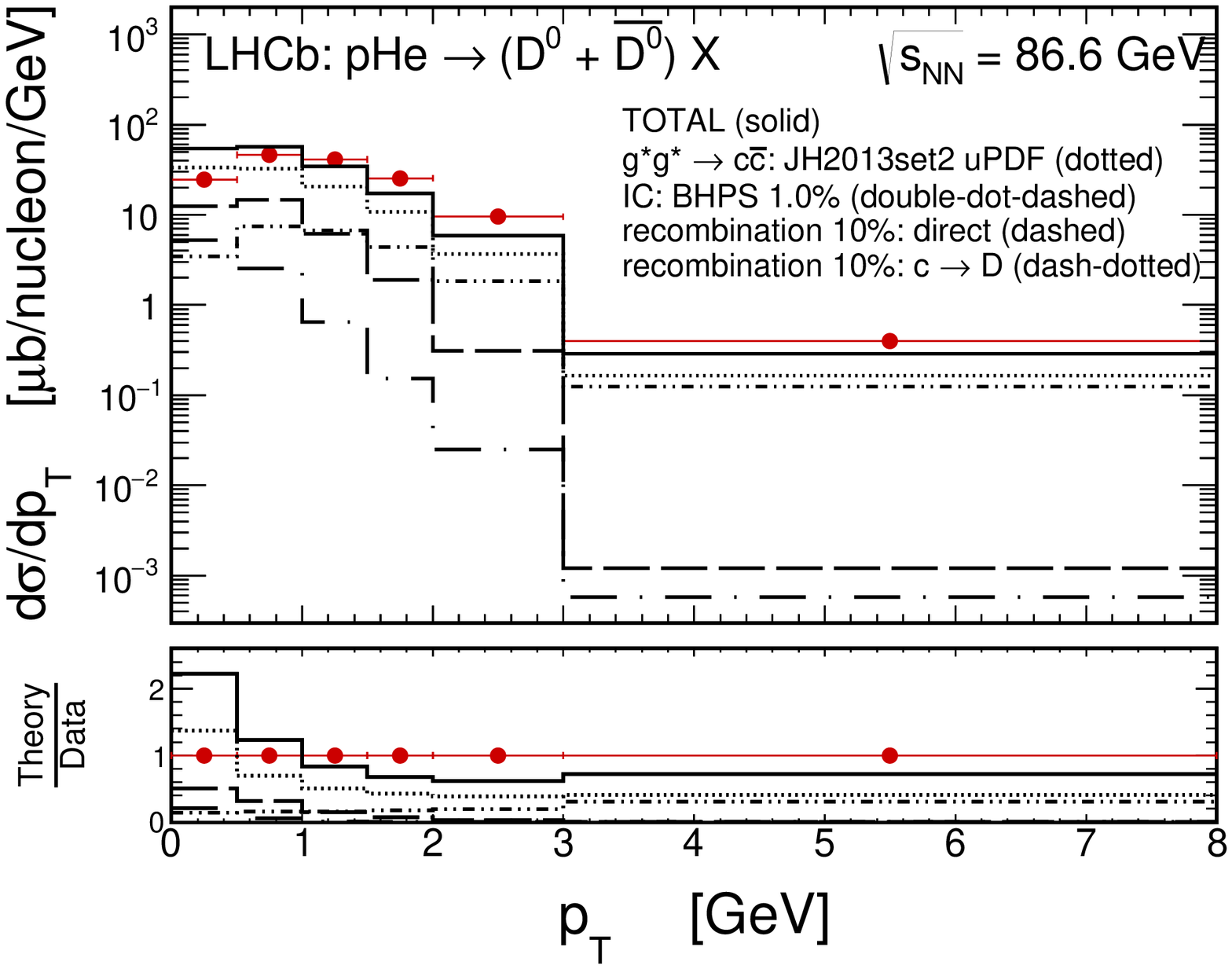}}
\end{minipage}
  \caption{
\small The rapidity (left) and transverse momentum (right) distributions of $D^{0}$ meson (plus $\overline{D^{0}}$ antimeson)
for $p+^{4}\!\!H\!e$ collisions at $\sqrt{s} = 86.6$ GeV together
with the LHCb data \cite{Aaij:2018ogq}. Here four different
contributions to charm meson production are are shown separately,
including the standard $g^*g^*\to c\bar c$ mechanism (dotted), 
the intrinsic charm contribution (double-dot-dashed) and two recombination 
components (dashed and dash-dotted). The solid histograms correspond 
to the sum of all considered mechanisms. Details are specified in the figure.
}
\label{fig:4}
\end{figure}

Let us now start presentation of numerical results with a discussion of
the LHCb fixed-target data on neutral open charm meson 
($D^0 + {\bar D}^0$)production in $p\!\!+\!\!^{4}\!He$ collisions at $\sqrt{s} = 86.6$ GeV. In Fig.~\ref{fig:4} we show the rapidity (left panel) and the transverse momentum (right panel)
 distributions for the $D^{0}$-meson. Here we plot separately four different components that lead to charm production: the standard $g^*g^* \to c \bar c$ mechanism (dotted), the intrinsic charm mechanism with the intrinsic charm probability $P_{IC} = 1\%$ (double-dot-dashed) and the two recombination contributions - the direct one (dashed) and the one from associated fragmentation (dash-dotted). The two former contributions are taken from our previous study of Ref.~\cite{Maciula:2021orz} and the recombination components are new. Here we assume the recombination probability $\rho = 10\%$. The solid histograms correspond to the sum of the four contributions. The predicted distributions are compared to the LHCb data taken from Ref.~\cite{Aaij:2018ogq}.
 
We observe that the recombination components may play an important role
in understanding the LHCb data in the region of backward
rapidities. There, according to our model the intrinsic charm
contribution is the dominant one, however, the recombination components
are only slightly smaller, depending of course on the recombination
probability parameter. Both recombination contributions are concentrated
only at small meson transverse momenta and are negligible at larger
$p_{T}$'s. On contrary, the intrinsic charm component seems to be
necessary to describe the large $p_{T}$ region. Therefore, in the
present analysis we shall assume that the presence of the recombination
components should not change the $P_{IC}$ parameter in order not to weaken the agreement between our model and the meson $p_{T}$ spectrum measured by the LHCb. In this scenario one could use the most backward rapidity bin of the distribution in meson rapidity to extract an upper limit for the recombination probability parameter $\rho$. For the intrinsic charm probability $P_{IC} = 1\%$ there is still a room for recombination contribution. Our study shows that the recombination component with the probability below 5\% seems to be not excluded by the LHCb data. Within our model its inclusion does not significantly weaken a very good description of the most backward LHCb data point where the cross section is dominated by the intrinsic charm component. Therefore $\rho=5\%$ for the recombination mechanism might be considered here as the rough estimation of its upper limit. The dependence of our predictions on the value of the $\rho$ parameter is shown explicitly in Fig.~\ref{fig:4}.  

\begin{figure}[!h]
\begin{minipage}{0.45\textwidth}
  \centerline{\includegraphics[width=1.0\textwidth]{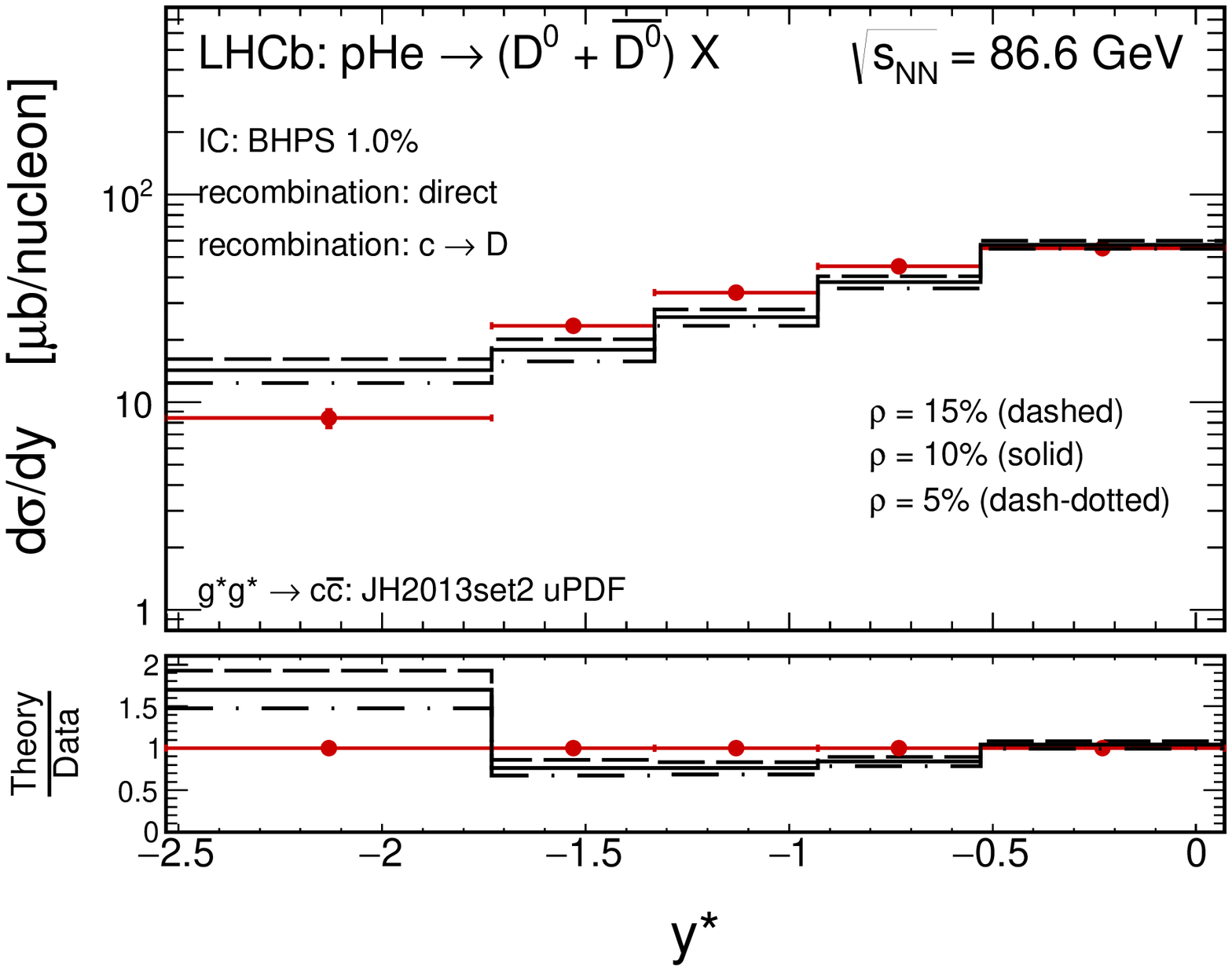}}
\end{minipage}
\begin{minipage}{0.45\textwidth}
  \centerline{\includegraphics[width=1.0\textwidth]{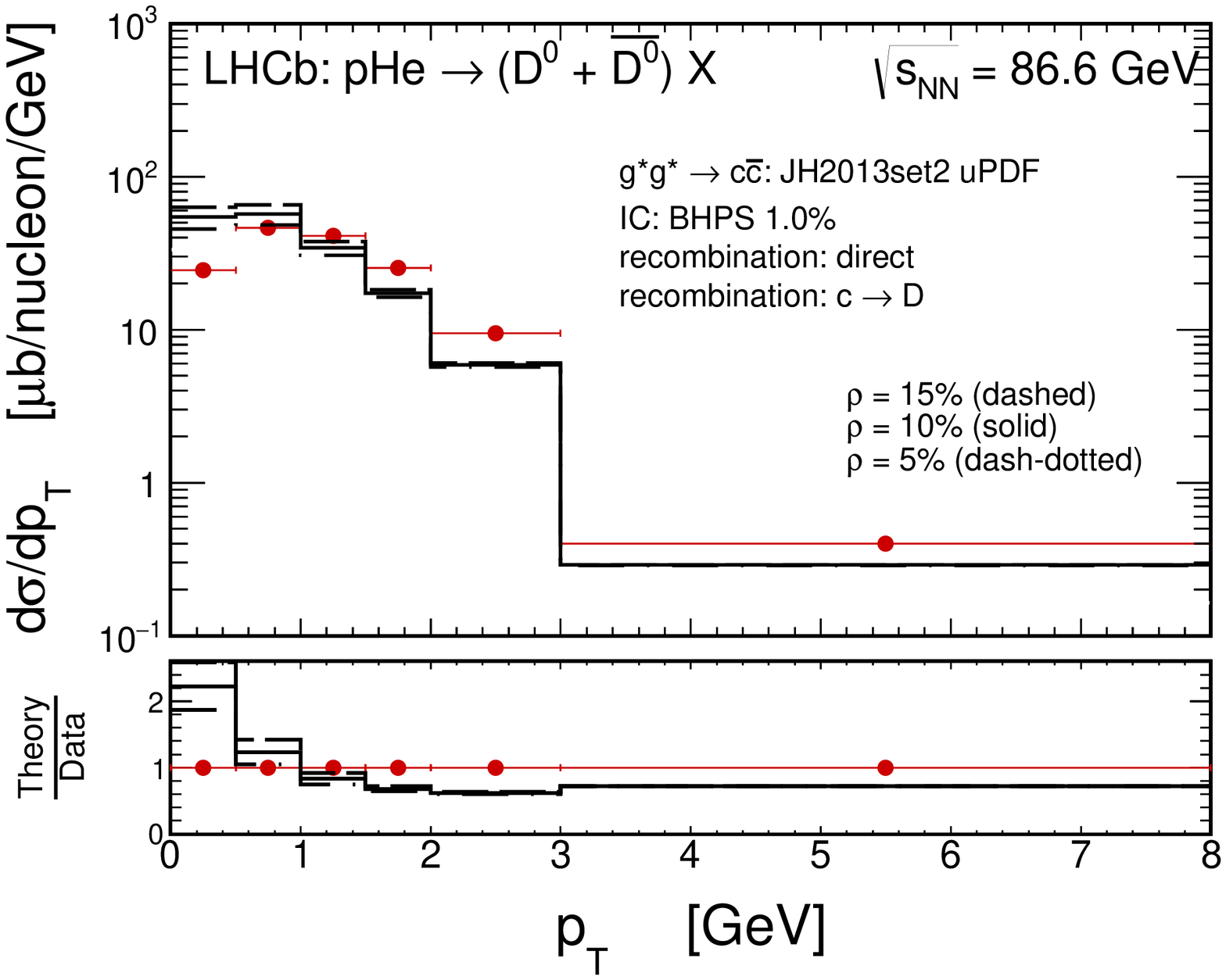}}
\end{minipage}
  \caption{
\small The same as in Fig.~\ref{fig:4} but here different histograms correspond to a total prediction of our model for three different values of the recombination probability $\rho = 15\%$ (dashed), $\rho = 10\%$ (solid), and $\rho = 5\%$ (dash-dotted). Details are specified in the figure.
}
\label{fig:5}
\end{figure}

\begin{figure}[!h]
\begin{minipage}{0.45\textwidth}
  \centerline{\includegraphics[width=1.0\textwidth]{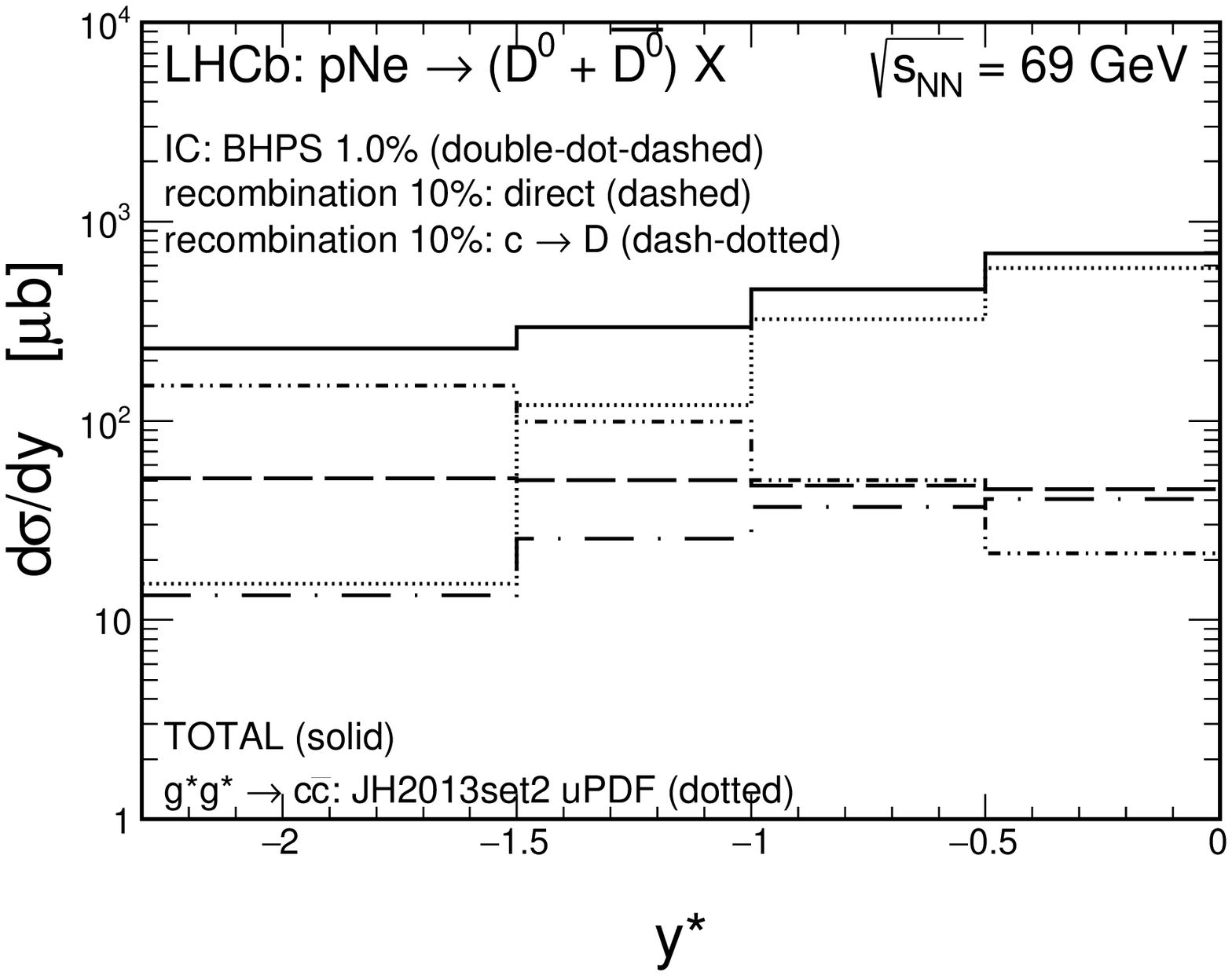}}
\end{minipage}
\begin{minipage}{0.45\textwidth}
  \centerline{\includegraphics[width=1.0\textwidth]{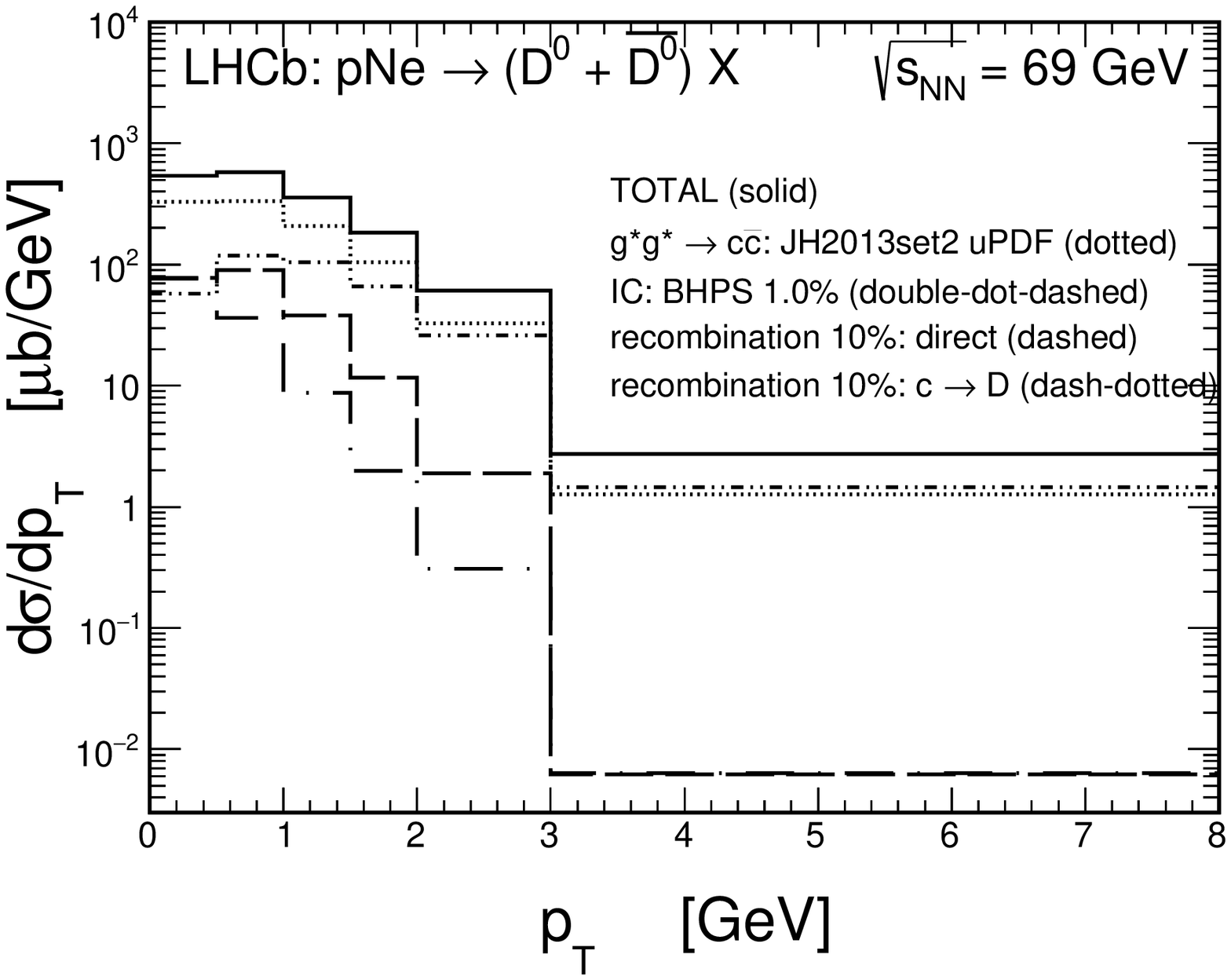}}
\end{minipage}
  \caption{
\small The rapidity (left) and transverse momentum (right) distributions of $D^{0}$ meson (plus $\overline{D^{0}}$ antimeson)
for $p+^{20}\!\!N\!e$ collisions at $\sqrt{s} = 69$ GeV within the
LHCb experimental acceptance. Here four different contributions to charm
meson production are shown separately, including the standard $g*g*\to
c\bar c$ mechanism 
(dotted), the intrinsic charm contribution (double-dot-dashed) and two
recombination components (dashed and dash-dotted). The solid histograms
correspond to the sum of all considered mechanisms. Details are 
specified in the figure.
}
\label{fig:6}
\end{figure}

In Fig.~\ref{fig:6} we show predictions of our model similar to those presented in Fig.~\ref{fig:4} but here we calculate the cross sections for $p+^{20}\!\!N\!e$ collisions at $\sqrt{s} = 69$ GeV within the LHCb experimental acceptance. However, the basic conclusions are the same as in the case of 
the $p+^{4}\!\!H\!e$ collisions at $\sqrt{s} = 86.6$ GeV discussed above.

\begin{figure}[!h]
\begin{minipage}{0.45\textwidth}
  \centerline{\includegraphics[width=1.0\textwidth]{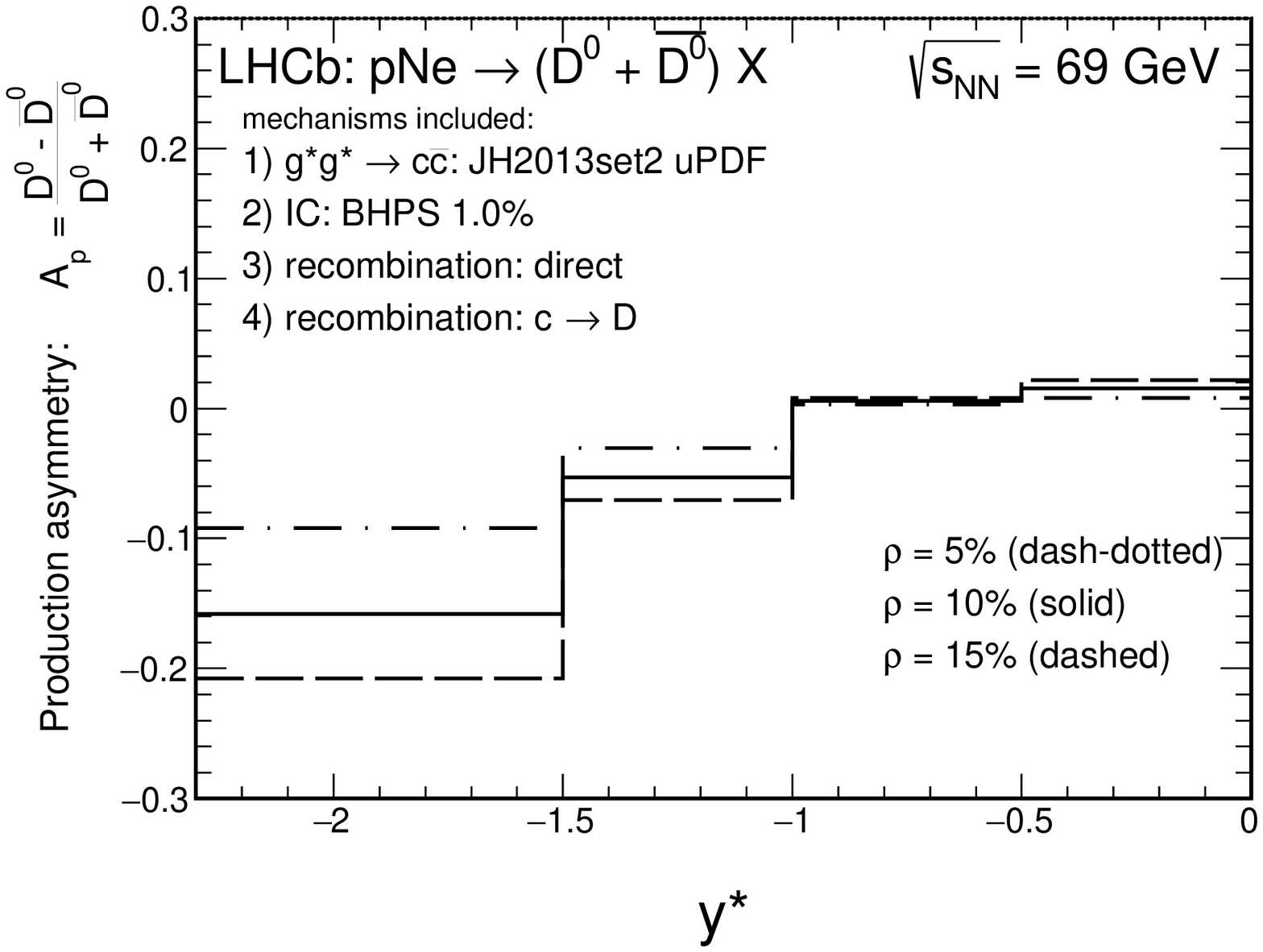}}
\end{minipage}
\begin{minipage}{0.45\textwidth}
  \centerline{\includegraphics[width=1.0\textwidth]{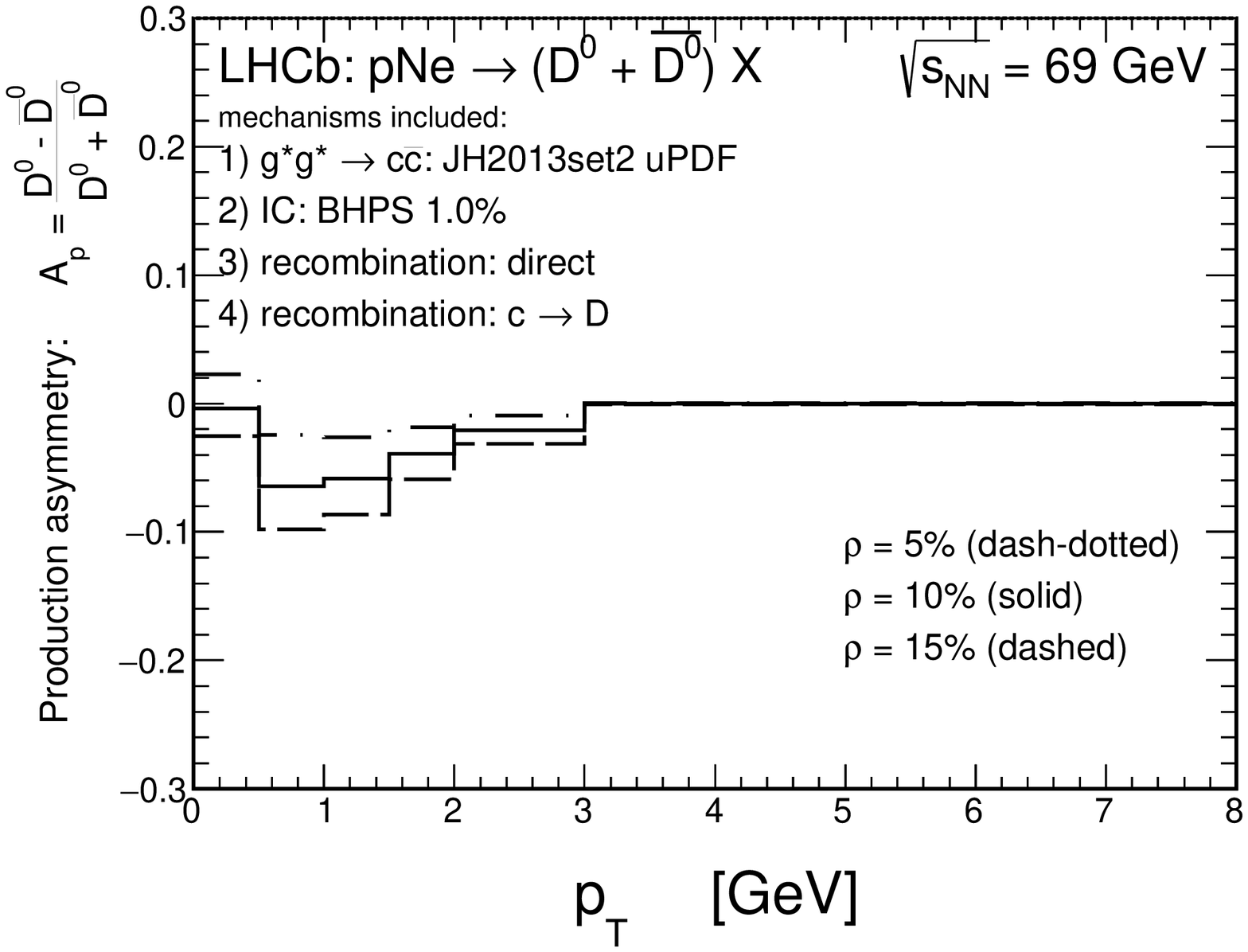}}
\end{minipage}
  \caption{
\small The production asymmetry $A_{p}$ for $D^{0}$-meson and $\overline{D}^{0}$-antimeson as a function of rapidity (left) and transverse momentum (right) for $p+^{20}\!\!N\!e$ collisions at $\sqrt{s} = 69$ GeV within the LHCb experimental acceptance. Here three different values of the recombination probability $\rho = 15\%$ (dashed), $\rho = 10\%$ (solid), and $\rho = 5\%$ (dash-dotted) are used. Details are specified in the figure.
}
\label{fig:7}
\end{figure}

Finally, we discuss production asymmetry for $D^{0}$-meson and $\overline{D}^{0}$-antimeson that can be also measured by the LHCb experiment.
The asymmetry is defined as: $A_{p} = \frac{D^{0} - \overline{D}^{0}}{D^{0} + \overline{D}^{0}}$. 
According to our present model only the recombination components may lead to the introduced production asymmetry. Therefore, we expect different values of the asymmetry for different values of the recombination parameter $\rho$. Future data for the asymmetry may therefore be also used to extract the probability for the recombination mechanism. The predictions of our model for the asymmetry $A_{p}$ as a function of meson/antimeson rapidity (left panel) and transverse momentum (right panel) are shown in Fig.~\ref{fig:7}. Here different histograms correspond to different values of the $\rho$ parameter. We get smaller production asymmetries for lower recombination probabilities $\rho$. The asymmetry depends on the meson  rapidity and transverse momentum. We predict the production asymmetry in the most backward rapidity bin probed by the LHCb fixed-target experiments as follows:
\begin{eqnarray}
 A_{p} = -0.096 \;\;\; \mathrm{for} \;\;\; \rho = 5\%, \\ \nonumber
 A_{p} = -0.160 \;\;\; \mathrm{for} \;\;\; \rho = 10\%, \\ \nonumber
 A_{p} = -0.208 \;\;\; \mathrm{for} \;\;\; \rho = 15\%, \nonumber
\end{eqnarray}
The predicted numbers can be confronted in future with the corresponding
LHCb measurement.

\begin{figure}[!h]
\begin{minipage}{0.45\textwidth}
  \centerline{\includegraphics[width=1.0\textwidth]{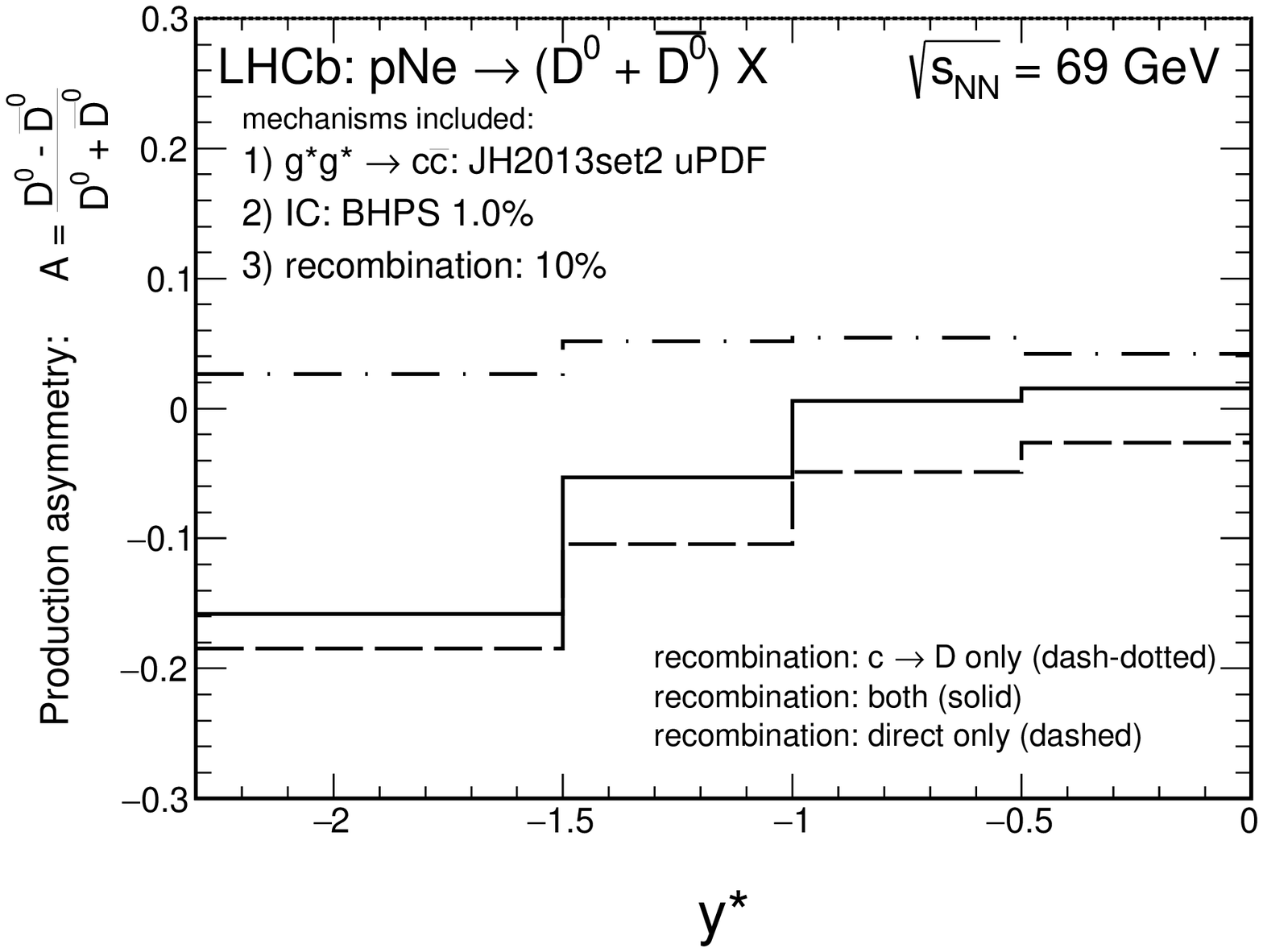}}
\end{minipage}
\begin{minipage}{0.45\textwidth}
  \centerline{\includegraphics[width=1.0\textwidth]{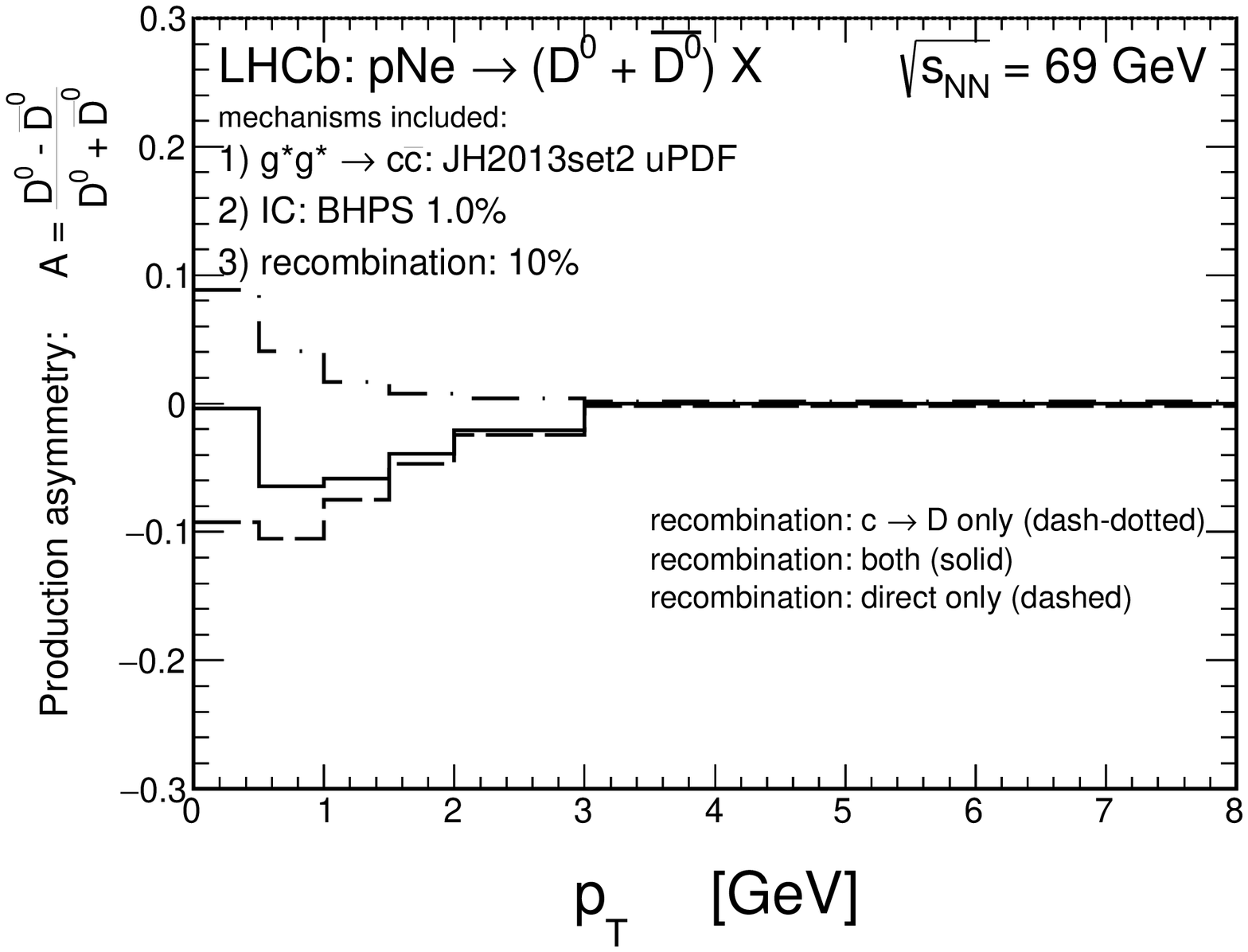}}
\end{minipage}
  \caption{
\small The production asymmetry $A_{p}$ for $D^{0}$-meson and $\overline{D}^{0}$-antimeson as a function of rapidity (left) and transverse momentum (right) for $p+^{20}\!\!N\!e$ collisions at $\sqrt{s} = 69$ GeV within the LHCb experimental acceptance. Here the recombination probability $\rho = 10\%$ is used. The asymmetry related with the direct only (dashed) and the $c \to D$ fragmentation only (dash-dotted) components of the recombination mechanism are shown separately. Details are specified in the figure.
}
\label{fig:8}
\end{figure}

The asymmetry we observed in the backward rapidity region is a negative number. It changes to a positive value when approaching midrapidities. The sign of the asymmetry is a result of interplay between the direct and the $c\to D$ fragmentation components of the recombination mechanism. The direct mechanism leads to a negative asymmetry while the $c\to D$ fragmentation component results in a positive one (see Fig.~\ref{fig:8}). The total asymmetry is the result of the cancellation between the two components and takes a negative sign in regions where the direct recombination component dominates over its fragmentation counterpart.

\section{Conclusions}

In the present paper we have discussed production of $D^0 + {\bar D}^0$
mesons in fixed target $p\!+\!\!^{4}\!H\!e$ and $p\!+\!\!^{20}\!N\!e$  experiments
at the LHC.
In addition to the previously included gluon-gluon fusion
and intrinsic charm contributions we have included also
recombination mechanism as proposed some time ago by BJM.
The latter contributions has characteristics similar to
that of the intrinsic charm.

Two contributions are included in the recombination mechanism:
the directly produced $D^0 + {\bar D}^0$ mesons as well as
$D^0 + {\bar D}^0$ produced in fragmentation of associated $c$
or ${\bar c}$. The latter has been done with the help of phenomenological
fragmentation functions. In our analysis we have used the Peterson
fragmentation functions with well known $\epsilon$ parameter.

The recombination model has one free parameter - probability of
the D meson production. This parameter can be in principle
fitted to experimental data.
Our fit shows that values of the order of 5 \% - 15 \% seems
relevant for the fit to the cross section data for the $p\!+\!\!^{4}\!H\!e$
scattering.

A better extraction of this parameter can be obtained from
the fit to $D^0 - {\bar D}^0$ asymmetry.
The recombination model leads to such an asymmetry.
We have discussed such an asymmetry for the $p\!+\!\!^{20}\!N\!e$  collisions.
A future experimental data \cite{maurice} could be used to extract the
$\rho$ parameter. This would better constrain also contributions
to the cross section.

\vskip+5mm
{\bf Acknowledgments}\

We are indebted to Emilie Maurice for discussion of the LHCb $p\!+\!\!^{20}\!N\!e$ experimental analysis in the context of charm production.
This study was supported by the Polish National Science Center grant UMO-2018/31/B/ST2/03537
and by the Center for Innovation and Transfer of Natural Sciences and Engineering Knowledge in Rzesz{\'o}w.


\end{document}